\documentclass[reprint,
 amsmath,amssymb,
 aps,
 pra,
]{revtex4-1}
\usepackage{amsmath}
\usepackage{blkarray}
\usepackage{braket}
\usepackage{graphicx}
\usepackage{dcolumn}
\usepackage{bm}
\usepackage[style=base]{caption}
\usepackage{subcaption}
\usepackage{stackengine}
\begin{document}

\preprint{APS/123-QED}

\title{Quantum correlations and entanglement in a Kitaev-type spin chain}\author{Vimalesh Kumar Vimal}
 \email{vimalkv@iitk.ac.in}
\author{V. Subrahmanyam}%
 \email{vmani@iitk.ac.in}
\affiliation{%
Department of Physics,  Indian Institute of Technology, Kanpur 208016, India}

\date{\today}

\begin{abstract}

The entanglement and quantum correlation measures have been investigated for the ground state of a spin chain with a Kitaev-type exchange interactions on alternating bonds, along with 
a transverse magnetic field. There is a macroscopic degeneracy in the ground state for zero magnetic field, implying a quantum critical point. But peculiarly in this model, the entanglement measures do not show any singular behaviour in the vicinity of the critical point, as seen in the transverse-Ising model ground state and related models.  We have investigated different correlation measures analytically, that have been used for many solvable spin systems to track a quantum critical point. We compute the pair concurrence measure of entanglement, the pair quantum discord to track the quantum correlations, and a global entanglement measure and a multi-species entanglement measure to investigate multi-party entanglement, both analytically and numerically.
 The nearest-neighbour concurrence shows a peak structure as a function of magnetic field, near the critical point, for various values of the ratio of interaction strengths, but its derivative does not show a singular behaviour close to the critical point.  A similar behaviour is shown by the quantum discord and the global entanglement.
 The multi-species entanglement shows the most-pronounced signature of the critical point, in its first-order derivative, though the entanglement and its derivatives have a smooth behaviour in the critical region.
 
\end{abstract}

\maketitle

\section{\label{sec:level1}Introduction}

Entanglement in quantum systems, being a fundamental property with no classical physics analog, has been at the center of research  interests  in many scientific disciplines. It is a key ingredient in quantum information processing protocols, the quantum superdense coding and the quantum key distribution. The quantum entanglement between two components of a many body system, encodes the quantum correlations, both diagonal and off-diagonal quantum correlations, in a very non-intuitive way. Several different measures of entanglement have been proposed and studied extensively. Entanglement between two parts of a composite system can be generated and manipulated by controlling the interaction of its subsystems. To this end, many interacting spin models have been studied from the entanglement viewpoint. In particular, the
entanglement study of spin systems is dominated by exactly soluble one-dimensional quantum spin systems,  the Heisenberg spin systems and the XY spin chain with transverse magnetic field. 

In this paper, we will focus on the entanglement structure of a Kitaev-type spin chain with a transverse magnetic field. The original Kitaev model\cite{kitaev} for the honeycomb lattice, studied in the context of fault-tolerant quantum computation, is a rare example of two-dimensional solvable model, through Jordan-Wigner transformation. The ground state exhibits many exotic features of topological order, non-Abelian statistics and entanglement.  A one-dimensional version has been studied exploring the anyon excitations, the entanglement entropy, and the quench dynamics\cite{abhinav,subra,uma}. These spin models can be realized in systems of cold atoms and superconducting quantum circuits\cite{duan,you}. 
Here, we will study many different measures of quantum correlations, the entanglement and the quantum information that can be analytically calculated for  the ground state of
a one-dimensional Kitaev spin chain.

The most commonly used measure of a two-party entanglement, the concurrence measure\cite{wootters1,wootters2} for a pair of qubits in arbitrary mixed state has been most extensively studied for various one spin models \cite{arnesen,su}, and it has been shown to exhibit singular behaviour near a quantum critical point of the phase transitions\cite{chen,osterloh}. The quantum discord\cite{ollivier,kundu,raoul} is a measure of quantum correlations, defined using the bipartite mutual information and the conditional entropy, in many-body quantum system.  
Both the concurrence and the discord are quantum correlation measures, however, the concurrence is zero for  a separable quantum bipartite state, while the discord may not vanish. These correlation measures require computations of an order of $d^2$ observables, multi-spin diagonal and off-diagonal correlation functions,  if the state lies in a $d$ dimensional Hilbert space. For a subsystem with more than two qubits, it is difficult to calculate all the correlation functions and the entanglement measures.

Global entanglement is a measure\cite{meyer,brennen} of entanglement that captures the distribution of entanglement over various parts of the system, has been studied for the ground state and the time-evolved states in spin chains. This is an easily computable measure as it involves only single-qubit reduced density matrices, which do not contain direct information of pairwise correlations, nevertheless, it does capture the global aspects of the entanglement distribution by the way of averaging over all the qubits. Another contrary measure of entanglement is
the multi-species entanglement\cite{subra2}, that
captures the correlations between two species of the system spread and moving over the whole system, as opposed to the above measures that involve two different spatial parts of the system. This measure can also track the quantum critical point and the associated singularity.

In this paper, we will present analytical and numerical results of various quantum correlation measures for the one-dimensional Kitaev spin chain. In Sec. II, we will describe the Kitaev
model for a finite spin chain with $N$ sites, and its exact solution for the ground state\cite{abhinav,subra} using the Jordan-Wigner transformation. The ground state will be written as a direct product over $N/4$ momentum $q$ modes, each mode consisting of four momentum states with four momentum values $q-\pi, -q, q,\pi-q$.  We will compute the matrix elements of single-qubit and
two-qubit reduced density matrices, in terms of the local magnetisation and pair diagonal and off-diagonal correlation functions. These reduced density matrices will be used for discussing
pairwise quantum correlation measures, the concurrence measure in Sec. III, and the conditional entropy and the discord in Sec. IV. The multi-party quantum correlation measures, the global entanglement and the multi-species entanglement, will be discussed in Sec. V and VI, followed by Sec. VII containing the conclusion of the results.

\section{\label{sec:level2}Ground State of Kitaev 1D Spin Chain}

We consider  a one-dimensional spin chain of $N$,  an even number,  spins with a Kitaev-type interaction involving x-component of the spins nearest-neighbour pairs on odd-numbered pair, and an interaction involving y-component of the spins for the even-numbered pairs, along with a homogeneous external magnetic field. Thus nearest neighbour spins have  $x-x$ and $y-y$ exchange interaction for alternate pairs. This model is a simplification of the honeycomb-lattice Kitaev model in 2 dimensions\cite{kitaev}. The Hamiltonian for the model is written as 
\begin{equation}
H = J_x\sum\limits_{i=odd}^{N}\sigma_{i}^x\sigma_{i+1}^x + J_y\sum\limits_{i=even}^N\sigma_i^y\sigma_{i+1}^y + h\sum\limits_{i=1}^N\sigma_i^z .
\end{equation} 
where $j_x$ and $j_y$ are the interaction strengths at odd and even sights with its nearest neighbour respectively, and $h$ is global magnetic field strength transverse to interacting spin polarization. This model is different from the $XY$ spin chain that contains different  $x-x$ and $y-y$ interaction strengths for every pair, whereas, here, alternate pairs have $x-x$ or $y-y$ interaction strength. The key difference from the XY model is that  the ground state of this Kitaev spin chain has a macroscopic degeneracy in zero magnetic field, implying a long-ranged spin correlations for a nonzero field. This degeneracy is lifted  by applying a uniform transverse magnetic field or a staggered transverse field. The Hamiltonian is solved using exact diagonalization steps and eigenstates can be written as direct product of fermion modes in momentum space. The considered model has $N/4$ uncoupled modes, with $0<q<\pi/2$. Each mode has four momentum values $q-\pi, q, -q, \pi-q$, while in $XY$ spin model, there are $N/2$ uncoupled modes $q>0$, and each mode has mixing of two momentum values, $\pm q$. \\
The Hamiltonian can be diagonalised by  a series of transformations, the Jordan-Wigner and the Fourier transformations followed by a Bogoliubov transformation. The spin operators of the Hamiltonian are first transformed into fermion operators through a Jordan-Wignar transformation, given by
\begin{equation}
\sigma_i^\dagger = e^{i\pi \Sigma_{j=1}^{i-1}n_j}c_i^\dagger , \sigma^z= 2c_i^\dagger c_i-1.
\end{equation}
Here $c_i^+$ and $c_i $ are fermion creation and annihilation operators on $it$'h site, $n_j$ are number operators at $j$'th site. The spin states with $\sigma_i^z=\pm 1$ is mapped to occupied($n_i=1$) and unoccupied($n_i=0$) fermion states. 
The Hamiltonian can be further simplified by transforming fermion operators in the momentum space, given by
\begin{equation}
c_l = \frac{1}{\sqrt{N}} \Sigma_qe^{iql}c_q.
\end{equation}
The allowed value of $q$ runs over all possible momenta in first Brilliuoin zone. For even number of total spins $N$ and using periodic boundary conditions give possible values of momenta as $q=\pm\frac{\pi}{N}(1,3,5.....,N-1)$.
The Hamiltonian now can be simplified further and expressed in terms of decoupled modes, using four momentum values $q-\pi$, $-q$, $q$, $\pi-q$ for each uncoupled mode $q$. For each mode the 4-dimensional Hamiltonian matrix can be diagonalised for its eigenvalues and eigenvectors. Following\cite{subra}, the Hamiltonian can be written in a simple diagonal form in terms of
$\xi_{\eta_1,\eta_2}$ operators which are linear combinations of the four fermion operators in momentum space; we have
\begin{equation}
H= 2\sum_{q=0}^{\pi/2} \sum_{\eta_1,\eta_2=\pm1} \lambda_{\eta_1,\eta_2} \xi_{\eta_1,\eta_2}^{\dag}\xi_{\eta_1,\eta_2}.
\end{equation}
Here, the single-particle mode eigenvalue $\lambda_{\eta_1,\eta_2}$ is given by
\begin{equation} 
\lambda_{\eta_1,\eta_2}= \eta_1 |\epsilon_q | + \eta_2 \sqrt{ h^2+ |\epsilon_q|^2} ,
\end{equation} where $2 \epsilon_q \equiv J_x  e^{-iq}+J_y e^{iq}$.

 The lowest energy state can be easily constructed  from the vacuum state (that has no $\xi$ particle occupation), by occupying all the negative-energy modes (corresponding to the eigenvalues $\lambda_{-,-},\lambda_{+,-}$ for  $0<q<\pi/2$),  as $ |GS \rangle = \prod_{0<q<\pi/2} \xi_{-,-}^{\dag} \xi_{+,-}^{\dag}|vac\rangle$. 
 Using the momentum mode operators defined (for $0<q<\pi$) as, $\sqrt{2}F_\pm=  c_{q-\pi}\pm c_{-q}^{\dag}$, and $\sqrt{2}G_\pm=  c_{q}\pm c_{\pi-q}^{\dag}$,  and using the parametrisation 
 $ Re~ \epsilon_{q}=|\epsilon_1| \cos{\theta_q}, Im ~\epsilon_{q}=|\epsilon_q|\sin{\theta_q}$, the mode operators occurring in the ground state are given by,
 \begin{eqnarray}
  \xi_{-, -}^{\dag} &= {\cal C}~ F_+^{\dag} + i {\cal S}~ G_+^{\dag} +  h_1 {\cal C} ~F_-^{\dag} + i h_1 {\cal S} ~G_-^{\dag},\cr
   \xi_{+,-}^{\dag}&={\cal S} ~F_+^{\dag}- i {\cal C} ~G_+^{\dag}+  h_2 {\cal S}~ F_-^{\dag} -i h_2 {\cal C}~G_-^{\dag}.
  \end{eqnarray}
  In the above, we have used the parametrisation ${\cal C}=\cos{\theta_q}/2,{\cal S}= \sin{\theta_q}/2, h_1=h/\lambda_{-,-}, h_2=h/\lambda_{+,-}$.
 The ground state can be written in a more convenient form using the
 number states of the four momentum fermion operators for every mode. We will use for $q$, the number states $|n_{q-\pi},n_{-q}, n_q, n_{\pi-q}\rangle$, where the number operator eigenvalue takes $n_i=0,1$ for every momentum value. Using the momentum-mode number-state representation, the ground state can be written as
\begin{equation}
\begin{split}
 \ket{GS}= \prod_{0\leq q \leq \pi/2}^{} {\large [} \alpha\Ket{1001}+ \beta\Ket{1111}+ \gamma\Ket{1100}+\\(1-\beta) \Ket{0000}-\gamma\Ket{0011}+ \alpha\Ket{0110} {\large ]}.    
 \end{split}
\end{equation}
Here the probability amplitude coefficients are given by
 \begin{equation}
 \begin{split}
 &\alpha= \frac{-i(h_2-h_1)\sin\theta_q}{2(h_1+h_2)}, \quad   
  \beta= \frac{(1+h_1)(1+h_2)}{2(h_1+h_2)}, \\ \\
 &\gamma= -i\alpha\cot\theta_q,
 \end{split}
\end{equation}
where the parameters are given by,  
\begin{equation}
\begin{split}
& e = \frac{1}{2}\sqrt{((j_x+j_y)\cos q)^2+((j_x-j_y)\sin q)^2},\\
& h_{1} = \frac{h}{\-(e+\sqrt{e^2+h^2})},\quad
 h_{2} = \frac{h}{\-(e- \sqrt{e^2+h^2})},\\
& \theta_q = \sin^{-1} {\frac{(1-r)\sin q}{\sqrt{((1+r)\cos q)^2+((1-r)\sin q)^2}}},\\
& r = \frac{j_y}{j_x} .
\end{split}
\end{equation}

The ground state written in the above form is a direct product over all modes $q$, and for each mode only even occupation states occur. There is a macroscopic degeneracy in the
ground state of $2^{N/2-1}$ for $h=0$, and for a nonzero $h$ the ground state is non-degenerate, for all value of $J_x, J_y$, implying a quantum critical point. We will now construct the
various reduced density matrices, by performing partial traces over the ground state, that are needed for studying quantum correlations, discussed in the following sections. We briefly describe the computation of matrix elements of reduced density matrices in terms of the spin operators and also using the fermion operators below. The single-qubit reduced density matrix for the spin $i$ is defined as $ \rho_i= Tr^{\prime} |GS\rangle \langle GS|$, where the prime over the trace indicates a sum over the states for all spin excepting $i$'th spin. Similarly, the
two-qubit reduced density matrix for spins $i$ and $j$ is given by $\rho_{i,j}=Tr^{\prime} |GS\rangle\langle GS|$, the partial trace is over all states of all the spins  except the indicated $i$'th and $j'$the spins. The cyclic boundary conditions imply that all the single-qubit reduced density matrices are the same, $\rho_i=\rho_1$, and there are two inequivalent two-qubit nearest neighbour pair reduced density matrices, $\rho_{1,2}$ and $\rho_{2,3}$, depending on the pair is even or odd. Further, owing to the fact that the Hamiltonian commutes with $(-1)^{N_F}$,
where $N_F$ is the number of fermions in the state, even and odd numbered fermion states do not mix. This implies that only some off-diagonal elements would be nonzero for the
reduced density matrix, corresponding to even-even or odd-odd matrix elements.

Using the basis of $\ket{0}$ and $\ket{1}$, $\sigma_1^z-$ diagonal basis for the first spin, the single-qubit reduced density matrix is given by
\begin{equation}
\begin{split}
 \\
\rho_i=\rho_1= 
 \begin{matrix}
     \\
    \\ 
  \end{matrix}
 \begin{bmatrix}
    1-n_1 & 0 \\
    0 & n_1\\ 
  \end{bmatrix}.
\end{split}
\end{equation}
Here, the diagonal matrix element is given as the ground state expectation value,  $n_1=\langle(1+\sigma_1^z)/2\rangle=\langle c_1^\dagger c_1\rangle $ is the expectation value of the number operator at the site in the ground state. The fermion operators can be transformed to the momentum state operators,  and calculated from the ground state given above, as
\begin{equation}
\langle c_1^\dagger c_1\rangle =\frac{1}{2}-\frac{2}{N}\sum\limits_q\frac{h}{\sqrt{|\epsilon_q|^2+h^2}}.
\end{equation}

The quantum pair correlations for the nearest neighbour pair in 1-dimensional spin chain can be studied by the reduced density matrix for that pair. In the spin chain all odd-even(even-odd) pairs have the same interaction strength $j_x(j_y)$. That is correlations for pairs (i,i+1) for i odd will be same as the pair (1,2), and similarly for all even $i$ pairs would be same as
the pair (2,3). It suffices to study correlations for pairs (1,2) and (2,3), which will reflect the behaviour for all odd-even and even-odd pairs respectively. 
The two-site reduced density matrix $\rho_{1,2}$, for the odd numbered pair,  has the form given below, the X state form, using the product $\sigma_1^z \sigma_2^z$ diagonal basis $|00\rangle,|01\rangle,|10\rangle,|11\rangle$ :
\begin{equation}
\begin{split}
\rho_{1,2}=\rho_{odd}=
  \begin{bmatrix}
    u_{odd} \quad & 0 & 0 & \quad  x_{odd} \\
    0 & \quad w_{1odd} & \quad   y_{odd} &0 \\
   0  & \quad  y^*_{odd} & \quad  w_{2odd} &0\\
     x^*_{odd} \quad  &0 &0& \quad v_{odd} \\ 
  \end{bmatrix}.
  \end{split}
\end{equation}
Similarly, the reduced density matrix $\rho_{2,3}=\rho_{even}$ has a X state form, except all the nonzero matrix elements will be denoted as $u_{even}$ and so on.
Many matrix elements in the above are zero due to the even-odd symmetry of the Hamiltonian as we discussed before.
Each of the matrix elements represents a particular correlation function of the spins interaction,  the diagonal element involving the diagonal correlation function and
 the off-diagonal matrix elements represent the off-diagonal correlations of the two spins. Using the spin operator language, the matrix elements are given by\\
$u_{odd}= \langle{1-\sigma_1^z\over 2}{1-\sigma_2^z\over 2}\rangle, ~~v_{odd}= \langle{1+\sigma_1^z\over 2}{1+\sigma_2^z\over 2}\rangle$,\\
$w_{1odd}= \langle{1-\sigma_1^z\over 2}{1+\sigma_2^z\over 2}\rangle, ~~w_{2odd}= \langle{1+\sigma_1^z\over 2}{1-\sigma_2^z\over 2}\rangle$,\\
$x_{odd}= \langle {\sigma_1^-\sigma_2^-} \rangle, ~~ y_{odd}= \langle{\sigma_1^-\sigma_2^+}\rangle $. \\
All these expectation values can be calculated in momentum space of operators. We have calculated and checked that for both even and odd cases, the off-diagonal matrix element
is zero, $y_{odd}=y_{even}=0$, thus, simplifying the computation of various correlations.
Off-diagonal elements when expressed in terms of fermions operators, they have a bilinear form. Each operator can have different mode value. So, it is possible to have two different mode
and momentum values involving $c_{q_1}$ and $c_{q2}$ or we can have different modes with same momentum involving $c_{-q}$ and $c_{\pi-q}$, and similarly other possible values. The former possibility is discarded as the ground state is comprised of either no occupation or even number of occupations for each possible mode. When operators of two different modes act on the state it will create an odd occupation configurations which will result in zero contribution for the expectation value. The off-diagonal matrix element can be calculated using the
fermion operators in the momentum space, we have
 This correlation function in momentum space is 
 \begin{equation}
x_{odd}=\langle c_1^{\dag}c_2^{\dag}\rangle = \frac{1}{N}\sum_{q_1,q_2}~ e^{-i(q_1+2q_2)} \langle c_{q_1}^{\dag}c_{q_2}^{\dag} \rangle 
\end{equation}
   
 \begin{figure*}[t]
 {(a) \hskip 6.5 cm (b)}\\
\begin{subfigure}{1.0\textwidth}
\includegraphics[width=0.49\linewidth, height=6.5cm]{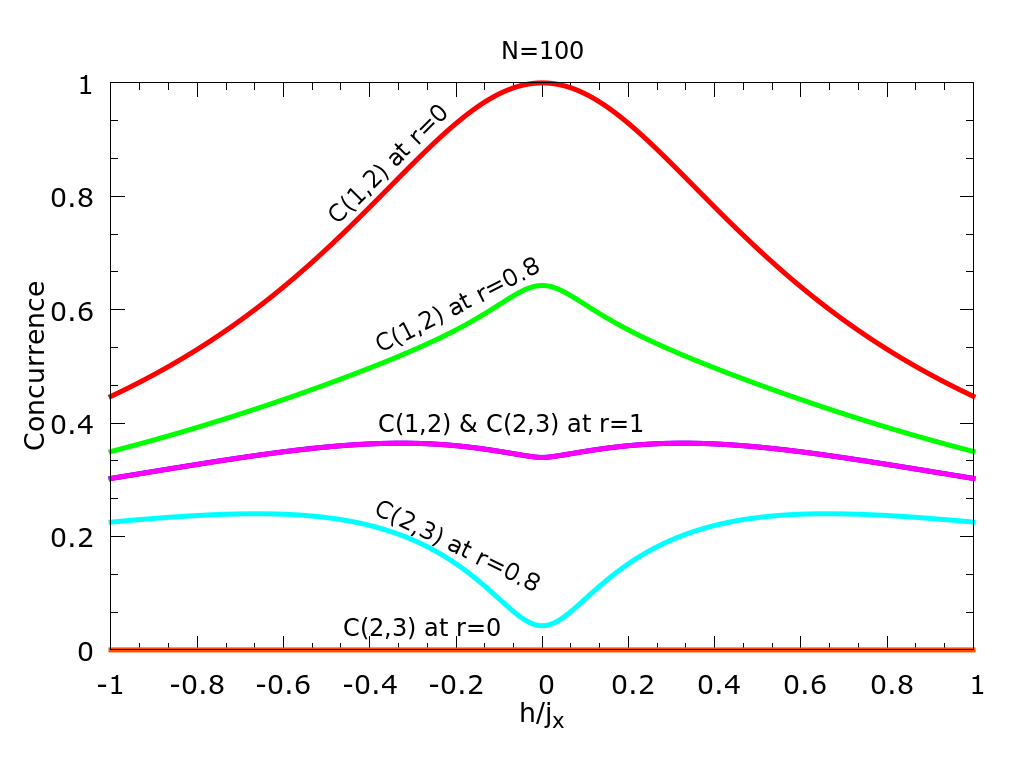} 
\includegraphics[width=0.49\linewidth, height=6.5cm]{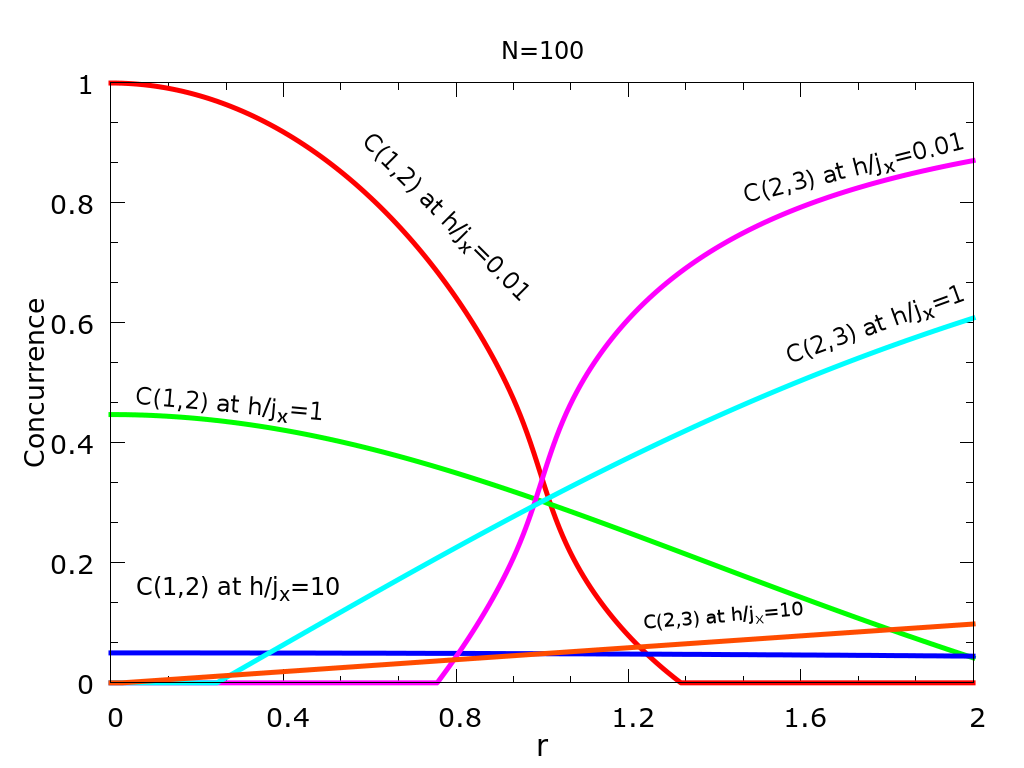}

\end{subfigure}
 \caption{(a) Concurrence of  $(1,2)$ and $(2,3)$ pairs of sites as a function of ratio of magnetic field and $xx$-interaction, $h/j_x$ at different values of ratios of strength parameters, $r$. At lower $r$-value , $C(2,3)$ decreases and the local minima gets more pronounced at $h=0$. For $r=0$, it follows the same pattern but comparatively, it is grounded. On the other hand, $C(1,2)$ increases with $r$ value, reaching a maximum at $h=0$, and becomes unity for $r=0$. (b) The comparative plots of even and odd concurrences as a function $r$, the ratio of the interaction strengths,  are shown at different $h/j_x$-values. $C(1,2)$  exhibits a cutoff behaviour, becomes zero for larger $r$, for lower $h/j_x$-values. $C(2,3)$ is zero for smaller values of $r$ for smaller $h$ values, and increases as $r$ increases. It reaches to its maximum at $r\rightarrow\infty$. $C(1,2)$ and $C(2,3)$ intersect at $r=1$ and these are almost negligible for $h=10$.  $j_x$ has been taken to be 1.}
\end{figure*}
 
The valid selection for $q_1$ and $q_2$  is given by different values in same mode, that is any two of $q-\pi$, $-q$, $q$, $\pi-q$ for a mode $q$. In the ground state, the contributing terms are 
 \begin{equation}
 \begin{split}
 &\langle c_{q-\pi}^+c_{-q}^+\rangle=\gamma(1-2\beta),~
 \langle c_{-q}^+c_{q}^+\rangle=\alpha(2\beta-1),\\
 &\langle c_{q-\pi}^+c_{\pi-q}^+\rangle=\langle c_{-q}^+c_{q}^+\rangle,~
 \langle c_{q}^+c_{\pi-q}^+\rangle=-\langle c_{q-\pi}^+c_{-q}^+\rangle.
 \end{split}
 \end{equation}
Using the above expectation values,  $x_{odd}$ can be calculated as,
\begin{equation}
x_{odd}=-\frac{2}{N}\sum\limits_{q<\frac{\pi}{2}}\frac{ |\epsilon_q| \cos(q-\theta_q)}{\sqrt{|\epsilon_q|^2+h^2}}.
\end{equation}
 Off-diagonal correlation functions are not difficult to calculate as the transformation of one spin operator into fermionic sector gives one fermion operator but along magnetic field the spin operator $\sigma^z$ corresponds to number operator in momentum space. So any diagonal element of density matrix involving two spin operators gives four fermion operators together in momentum space along with individual phases which when considered for selection of  modes and momenta, it posses great difficulty as it can have momenta in same or different modes. It is convenient to express diagonal matrix elements in terms of fermion operators. The diagonal element $w_{1odd}$ and $w_{2odd}$ are the same for our Hamiltonian so we write $w_{odd}=w_{1odd}=w_{2odd}$. The different diagonal elements are written in terms of the fermion operators as,
\begin{equation}
\begin{split}
w_{odd}=\langle c_1^\dagger c_1\rangle -\langle c_1^\dagger c_1c_2^\dagger c_2\rangle, \quad \quad \quad \quad \quad \quad \\
u_{odd}= 1+ \langle c_1^\dagger c_1c_2^\dagger c_2\rangle -2\langle c_1^\dagger c_1\rangle, \quad \quad \quad \quad \\
v_{odd}= \langle c_1^\dagger c_1c_2^\dagger c_2\rangle. \quad \quad \quad \quad \quad \quad \quad \quad \quad \quad  \\
\end{split}
\end{equation}
The $\langle c_1^\dagger c_1c_2^\dagger c_2\rangle $ term when transformed in momentum space it has four operators with different momentum values. These momentum may or may not lie in same mode which makes it tedious to calculate as the phase is also there for each momentum. Looking at different possibilities, the leading terms are
$ \langle c_1^\dagger c_1c_2^\dagger c_2\rangle \approx n_1^2 + |x_{odd}|^2$.  
  
For the even-numbered pair correlations, it suffices to study the matrix elements of $\rho_{2,3}$, with a similar structure shown in Eq. 12. The corresponding off-diagonal matrix element is, in analogy with Eq. 13, given by
 \begin{equation}
x_{even}=\langle c_2^{\dag}c_3^{\dag}\rangle = \frac{1}{N}\sum_{q_1,q_2}~ e^{-i(2q_1+3q_2)} \langle c_{q_1}^{\dag}c_{q_2}^{\dag} \rangle.
\end{equation}
Using the relations given in Eq. 14, the matrix element can be rewritten, analogous with Eq. 15, as
\begin{equation}
x_{even}=-\frac{2}{N}\sum\limits_{q<\frac{\pi}{2}}\frac{ |\epsilon_q| \cos(q+\theta_q )}{\sqrt{|\epsilon_q|^2+h^2}}.
\end{equation} 
That is $x_{even}$ can be obtained by changing the sign of $\theta_q$ in $x_{odd}$ and vice versa.  Similarly the diagonal matrix elements for the even case are given, in analogy with
Eq. 17, as
\begin{equation}
\begin{split}
w_{even}=\langle c_2^\dagger c_2 \rangle - \langle c_2^\dagger c_2 c_3^\dagger c_3\rangle, \quad \quad \quad \quad \quad \quad \\
u_{even}= 1+ \langle c_2^\dagger c_2 c_3^\dagger c_3\rangle -2\langle c_2^\dagger c_2\rangle, \quad \quad \quad \quad \\
v_{even}= \langle c_2^\dagger c_2 c_3^\dagger c_3 \rangle. \quad \quad \quad \quad \quad \quad \quad \quad \quad \quad  \\
\end{split}
\end{equation}
Similar to the odd case, the leading terms of the two-point density-density correlation function for the even case is given by,  $ \langle c_2^\dagger c_2 c_3^\dagger c_3\rangle \approx n_1^2 + |x_{even}|^2$.  In the following sections we will use the diagonal and off-diagonal matrix elements in computing the quantum entanglement and correlation measures.

\section{\label{sec:level3}Concurrence}

Quantum correlations exhibited by a non separable quantum state have been studied extensively for many spin systems. Some of the pairwise entanglement measures that have been
studied are the pair concurrence\cite{wootters1,wootters2}, the quantum discord\cite{ollivier,kundu,raoul}, the tangle and the entanglement negativity. In this section we are studying the concurrence measure of entanglement, 
$C(i,j)$, for $(i,j)$ pair of qubits in the many-qubit ground state of the Kitaev model we have described in the previous section. It turns out that only nearest-neighbour pair of spins
have a nonzero concurrence in the Kitaev model ground state for all parameter values, similar to $XY$ model in transverse magnetic field. That is, only $C(i,i+1)$ the nearest-neighbour
pair concurrences are nonzero. There are only two independent pair concurrences, correspond to the even and odd nearest-neighbour pair of spins, viz. $C(i,i+1)=C(1,2)$ for $i$ odd, and $C(i,i+1)=C(2,3)$ for $i$ even.

The concurrence for two qubits is calculated from the two-qubit reduced  density matrix $\rho$,  viz. from $\rho_{1,2}$ for odd pairs and from $\rho_{2,3}$ for even pairs. Using  Wootters' formula, the 
concurrence is given by,
 \begin{equation}
 C= 2 max(\lambda_1-\lambda_2-\lambda_3-\lambda_4, 0),
 \end{equation}
where $\lambda_i'$s are  eigenvalues arranged in decreasing order of the Hermitonian matrix given by  $R= \sqrt{\sqrt{\rho}\tilde{\rho}\sqrt{\rho}}$.  Here, $\tilde{\rho} = \sigma^y\otimes\sigma^y\rho~\sigma^y\otimes \sigma^y$, is the time-reversed density matrix of $\rho$ 
In our case,  the  concurrence $C_{odd}$ can be calculated using the diagonal and off-diagonal correlation function discussed in the previous section, as 
\begin{equation}
   C(1,2)= 2(|x_{odd}|-|w_{odd}|),
   \end{equation}
when the off-diagonal correlation function is larger than the diagonal function, $|x_{odd}|>|w_{odd}|$, otherwise the concurrence is zero.

\begin{figure*}[t]
 {(a) \hskip 6.5 cm (b)}\\
\begin{subfigure}{1.0\textwidth}
\includegraphics[width=0.49\linewidth, height=6.5cm]{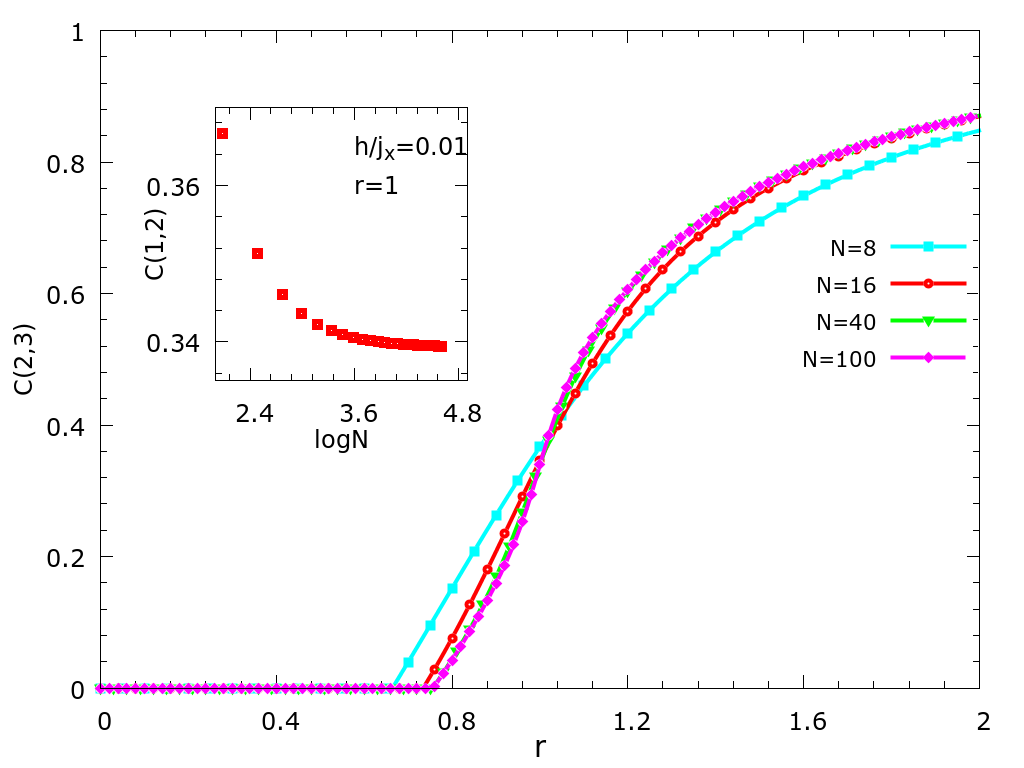}
\includegraphics[width=0.49\linewidth, height=6.5cm]{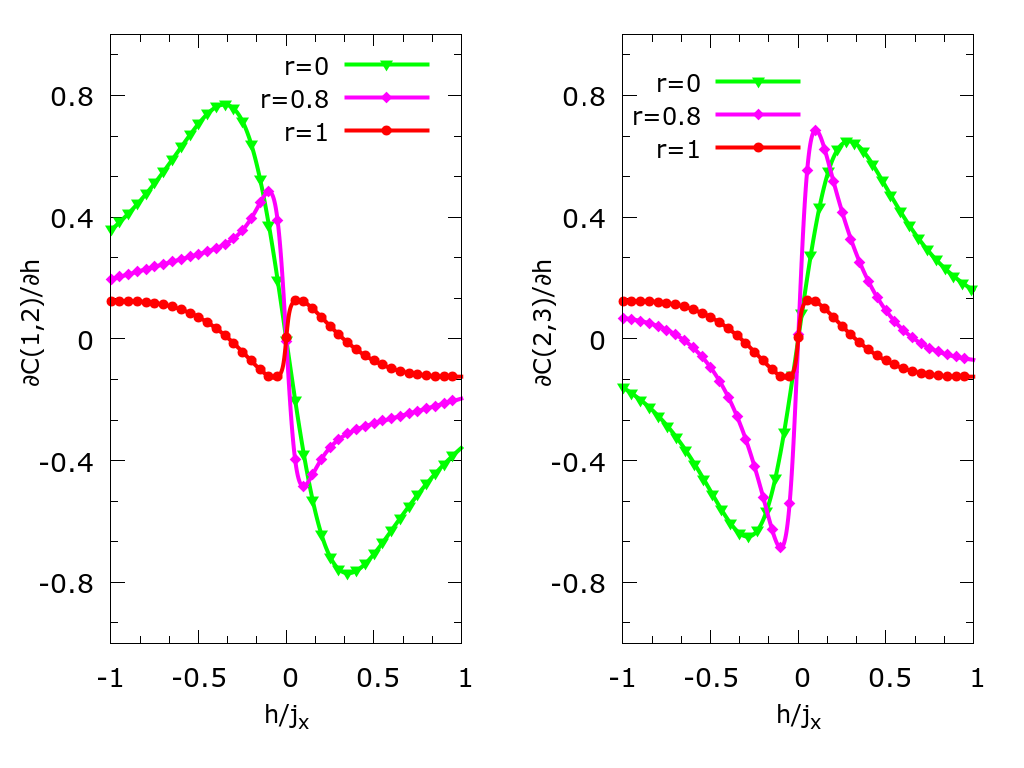}  

\end{subfigure}
 \caption{(a). Even concurrence as a function of $r$, for various $N$ values in zero magnetic field. For larger $r$ values, for a larger chain size, $C(2,3)$ increases up to $N=40$ above that it becomes constant. (b). The first-order derivatives of the concurrences for odd pair $(1,2)$ and even pair $ (1,2)$ are shown with respect to magnetic field at different $r$ values. There is
 no singular behaviour here in the critical region for an infinite chain, unlike other one dimensional spin models. }
\end{figure*}

Using Eq.15 and Eq.16, the leading terms in the concurrence, for large $N$, can be worked out explicitly.  The concurrence is given now as,
\begin{equation}
\begin{split}
C(1,2)= |x_{odd}|(1-|x_{odd}|)+n_1(1-n_1) + O(1/N),
\end{split}
\end{equation}
in terms of the off-diagonal correlation function and the number operator expectation value. A similar expression 

for $C(2,3)$ holds by replacing $x_{odd}$ by
$x_{even}$ in the above expression. The two pair concurrences differ by a change of the sign of $\theta_q$ in the off-diagonal correlation function, that we discussed in the last section.

In the thermodynamic limit $N\rightarrow\infty$ and $h=0$ case the above expressions for the concurrence can be simplified, by replacing the sums in Eq.11 and Eq.15 for the diagonal
and the off-diagonal matrix elements.  The concurrence is now expressed in terms of elliptical functions, as
 \begin{eqnarray}
C(1,2)&=  \frac{1}{4}+f(1-f),\\
C(2,3)&= \frac{1}{4}+g(1-g),
\end{eqnarray}
where, the two elliptic functions are defined as, 
\begin{eqnarray}
f&=\int_{0}^{\pi/2}\frac{dq}{\sqrt{1-x^2\sin^2q}}-\frac{2r}{1+r}\int_{0}^{\pi/2}\frac{dq\sin^2q}{\sqrt{1-x^2\sin^2q}},\\
g &=\int_{0}^{\pi/2}\frac{dq}{\sqrt{1-x^2\sin^2q}}+\frac{2r}{1+r}\int_{0}^{\pi/2}\frac{dq\sin^2q}{\sqrt{1-x^2\sin^2q}},
\end{eqnarray}
in terms of the ratio of the exchange interactions, $x\equiv 4r/(1+r)^2$. For nonzero magnetic field, no simplification of the concurrence in terms of elliptic functions as above, and we have to work with more complicated expressions for the concurrences using Eq.11, Eq.15 and Eq.22 directly.

We will discuss below the behaviour of pair concurrence as the interaction strength or the magnetic field is varied, and as function of the size of the spin chain. It is expected, similar to
the transverse-Ising model ground state, that the concurrence will exhibit singular behaviour close to macroscopic degeneracy point or the quantum critical point at $h=0$, for any value of the ratio of the interaction strengths\cite{you}.
Among the pairwise concurrences, only  nearest-neighbour concurrences, viz. the two distinct concurrences $C(1,2)$ and $C(2,3)$, are non for all values of the parameters, similar to the
case of the transverse-field Ising/XY model ground state. The presence of more than 4 operators with phase makes it a difficult task for analytical calculation for  next  neighbours concurrences $C(1,3)$ and $C(1,4)$, but the numerical analysis shows that these concurrences are not present in the system for any arbitrary magnetic field strength and relative interaction strength ratio. 
The nearest-neighbour pair concurrences are shown as functions of the ratio $h/j_x$ of the magnetic field and the interaction strength $j_x$  in Fig.1(a), for a chain of $N=100$ spins for different values of the ratio $r=j_y/j_x$ of the interaction strengths.  The odd concurrence $C(1,2)$ (even concurrence $C(2,3)$) exhibits a peak (dip) at $h=0$, the critical point. For $r=0$ (also for very large values), the spin chain is a set of $N/2$ disconnected nearest neighbour pairs, we have only $C(1,2)$ (only $C(2,3)$ nonzero for large $r$) nonzero.  For $r=1$ both concurrences merge, as both the interaction strengths are same implies that both
odd and even bonds are same. For $r>1$, not shown in the figure, the odd concurrence will exhibit a dip and the even concurrence showing a peak structure.
The behaviour of concurrence on either side of $h=0$ is same, which can be inferred from the Eq. (22),  as the expression for $C(1,2)$ is not altered by  changing the sign of the magnetic field. Furthermore, in region $0<r<1$ when $r$-value increases the maxima of concurrence reduces to half of its value at $r=0$. 

The pair concurrences are plotted as functions of the interaction strength for different magnetic field in Fig. 1(b) for a chain with $N=100$ spins.  Here, the odd  concurrence decreases continuously as function of the ratio of the strength, and becomes zero after a cutoff value of $r$. The  cutoff value of $r$ increases as the magnetic field strength increases.   The even concurrence becomes nonzero after a cutoff value of $r$,  and takes its maximum value as $r\rightarrow \infty$,  with the cutoff decreasing with increasing magnetic field. 
The crossing of the two pair concurrence plots at $r=1$ represents the fact that both the concurrences have equal value as explained above in Fig. 1(a).
The odd (even) concurrence becomes zero beyond  a cutoff value, due to the off-diagonal correlation function falling below a threshold value for dominating over the diagonal correlation function, as shown in Eq. 21, to give a nonzero concurrence. We will discuss this aspect of the off-diagonal correlation function as a function of $r$ and $h$ in in more detail below.  The off diagonal correlation dominates in lower $r$ region, which reflects in nonzero concurrence $C(1,2)$ in Fg.1(b).

\begin{figure*}
\includegraphics[width=0.5\linewidth, height=6.5cm]{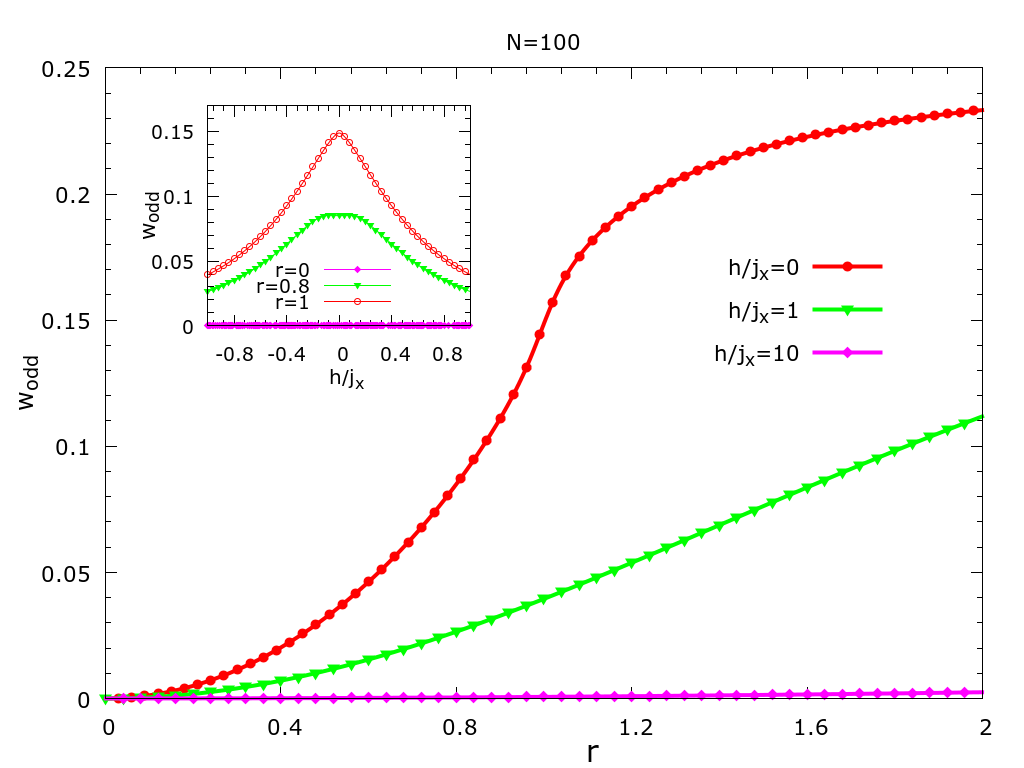}

\caption{The nearest neighbour correlation function, $W_{odd}$ as a function of $r$ at different $h/j_x$-values. The inset plot shows $W_{odd}$ as a function of $h/j_x$ at different $r$-values.  }
\end{figure*}

The finite-size effect is illustrated in Fig.2(a), showing the even concurrence $C(2,3)$ as a function of $r$ in zero magnetic field for different $N$-values. The cutoff value of $r$, beyond which the even concurrence is nonzero, increases by a small amount with the size $N$ of the system. The concurrence is nonzero after the cutoff, and approaches to its maximum value
for large value of $r$. The odd concurrence $C(1,2)$ decreases with an increase in the size of the chain, but converges quickly after around$N=40$, as the inset plot shows the concurrence as function of $\log(N)$ at $r=1$ and near zero magnetic field, in the vicinity of the critical point. This is in contrast to the critical behaviour of pairwise entanglement measure in the vicinity of a quantum phase transition in different spin models, that have been widely investigate\cite{su,chen,osterloh}. In the 1D Heisenberg spin chain it has been shown that the pairwise entanglement reaches to its local maxima at quantum phase transition point. In the Ising model, the pure ground state entanglement by entropy measurement shows that entanglement is increasing as function of the critical parameter. We see a reminiscent behaviour of the pair concurrences in Fig.2(b),  showing the first order derivative of the concurrence with the magnetic field. In sharp contrast to the transverse-Ising model ground state, the derivative of the concurrence does not diverge in our case at the critical point. Though it exhibits a sharp  feature at
the critical point of $h=0$,  here, the first order derivative of concurrences for the Kitaev type chain exhibits a smooth behaviour in the critical region, as seen in Fig. 2(b). 
This smooth behaviour continues for at all parameter values, and the pair concurrence  shows no diverging signature in the critical region, even for an infinite chain.
As we have stated above, the concurrence behaviour as a function of the interaction strength and the magnetic field can be understood from the behaviour of the diagonal correlation function. In Fig.3, we have shown the diagonal correlation function $w_{odd}$ as a function of $r$ for a few values of the magnetic field. The behaviour is similar to the odd concurrence 
shown in Fig.1(b), except that there is no cutoff behaviour for the correlation function.  We see here that the curve for $h=0$ is above
the curve for a nonzero field, implying the a peak in the odd concurrence at $h=0$. The off-diagonal correlation function does not show any cutoff value for it be nonzero, but it will dominate over the diagonal correlation function beyond a cutoff, translating into a cutoff for the odd concurrence. The inset in Fig.3 shows the diagonal correlation as a function of $h$ for a few
values of $r$, the behaviour is similar to that of the odd concurrence seen in Fig. 1(a), displaying a peak structure at $h=0$. For $r=0$, the diagonal correlation function goes to zero, which makes the concurrence value to be double of the value of the off-diagonal correlation function. We will correlate the behaviour of the diagonal correlation function further with other measures of quantum correlations in the next sections.
 
 \section{\label{sec:level4}Quantum Discord}
  
 The concurrence measure we have studied in the last section gives an insight into the entanglement structure of the state, but
it fails to detect the existence of quantum correlations. In other words, it is nonzero if the off-diagonal order dominates over the diagonal correlations (see Eq. 21). 
A state with zero concurrence still can exhibit quantum correlations. Quantum discord is a measure of quantum correlation \cite{ ollivier,kundu,raoul} that can address the quantum information present is separable as well as non-separable quantum state. This measure is defined using the mutual information and marginal subsystem entropies, and the mutual information and the conditional entropy; and it uses the mismatch of the two classically equivalent ways of defining the mutual information. A conditional entropy of a bipartite quantum state involves a measurement on one of the two parts comprising the composite state, and the mutual information thus defined may not match with the mutual information defined using the subsystem entropies. 
 
Let us consider a two-qubit  $\rho_{12}$,  of the two spins 1 and 2. The state $\rho_{1(2)}$ is the reduced density matrix of site 1(2), obtained by a partial trace over site 2 (1). The mutual information shared between two sites 1 and 2 has two different expressions which are same in classical case but differ for the quantum state. 
 We can define the mutual information $J(\rho_{12})$ in terms of von Neumann entropies of the subsystems as,
 \begin{equation}
J(\rho_{12}) = S(\rho_1)+S(\rho_2)-S(\rho_{12}),
 \end{equation}
 
\begin{figure*}
\includegraphics[width=0.5\linewidth, height=6.5cm]{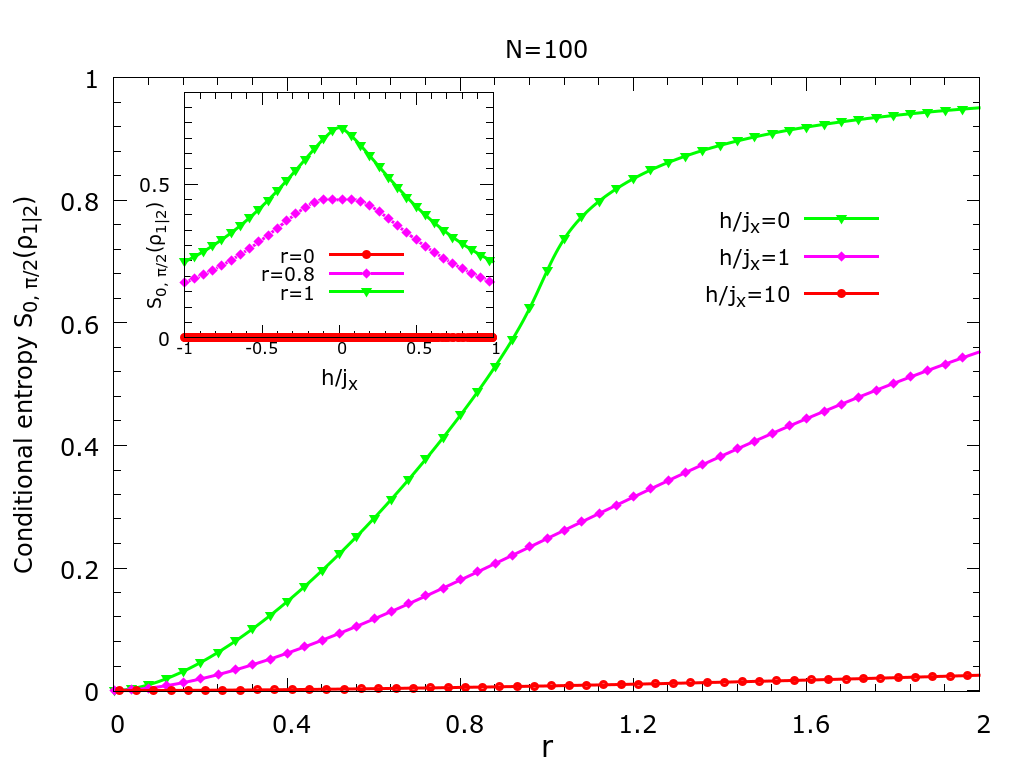}

\caption{The minimum conditional entropy as a function of $r$ at different $h$-values. The inset plot shows $\mathcal{S}_{0,\frac{\pi}{2}}(\rho_{1|2})$ as a function of $\frac{h}{j_x}$ at different $r$-values. The behaviour seen here is similar to the correlation function in Fig.3 }
\end{figure*}

 where $ S(\rho) = -Tr(\rho\log\rho)$ is Von Neumann entropy of density matrix $\rho$. Another way the mutual information can be defined using the conditional entropy as,
 $I(\rho_{12})= S(\rho_{1})- \mathcal{S}(\rho_{1|2})$. Now, $\mathcal{S}(\rho_{1|2})$ is the conditional entropy of qubit 1 when the qubit 2 has been measured. Let us set a measurement basis for qubit 2 as $\ket{\tilde 0}$ and $\ket{\tilde 1}$,  defined in terms of the eigenstates of $\sigma_2^z$ operator, as

\begin{equation}
\begin{pmatrix}
 \ket{\tilde{0}} \\
  \ket{\tilde{1}}\\
\end{pmatrix}
=
  \begin{pmatrix}

    \cos \theta'/2 & e^{i\phi^\prime} \sin \theta'/2 \\
    \sin \theta'/2 & -e^{i\phi^\prime} \cos \theta'/2 \\  
  \end{pmatrix}
  \begin{pmatrix}
  \ket{0} \\
  \ket{1}
  \end{pmatrix}.
\end{equation}
In the above we have characterised the measurement basis with the angles $\theta',\phi'$, which can be varied to find the minimum difference between the two mutual information
quantities we have defined. 

 \begin{figure*}[t]
 {(a) \hskip 6.5 cm (b)}\\
\begin{subfigure}{1.0\textwidth}
\includegraphics[width=0.49\linewidth, height=6.5cm]{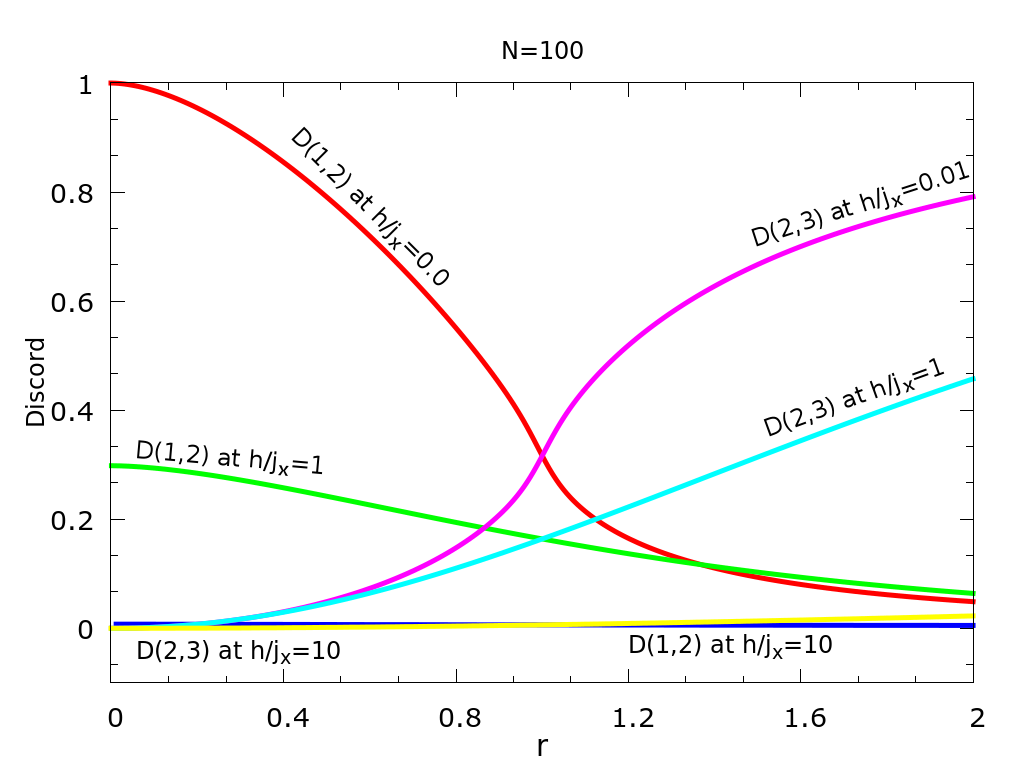} 
 \includegraphics[width=0.49\linewidth, height=6.5cm]{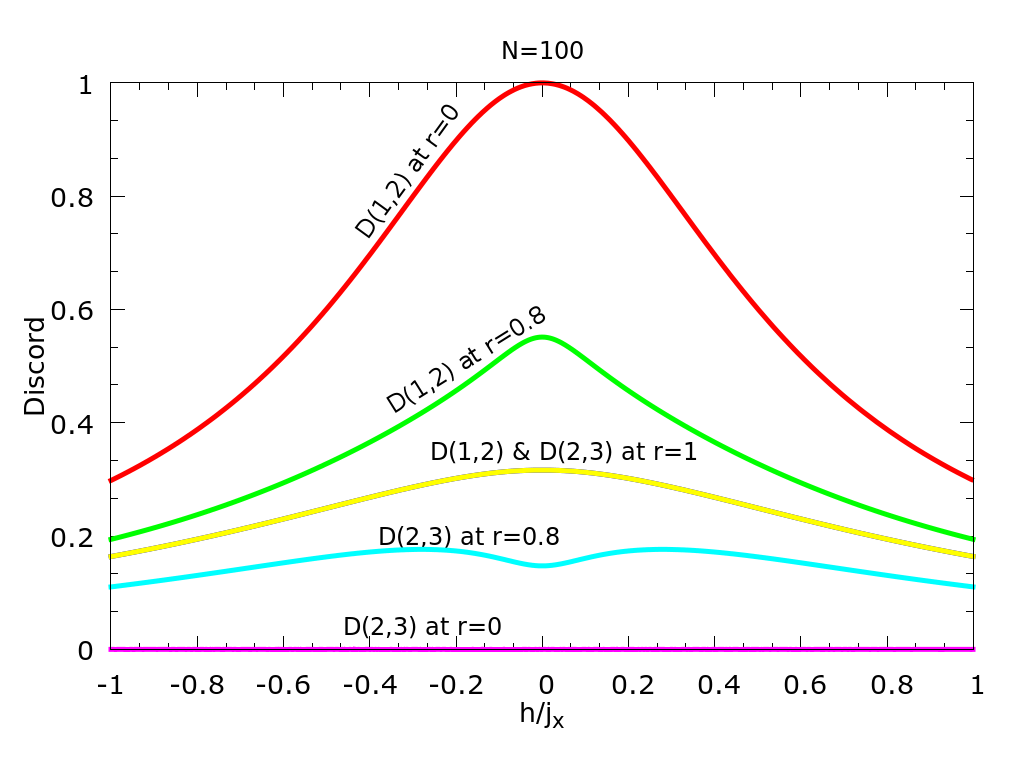}
\end{subfigure}
 
\caption{(a). Discord for the odd pair $(1,2)$ and the even pair $(2,3)$ as a function of $r$ at different $h/j_x$-values. The discord is minimised  for $\phi'=0$ and $ \theta'= \pi/2 $. The mutual information for both pairs show a similar pattern. (b). The discord for the odd pair $(1,2)$ shows a  peak at $h=0$ and approaches unity for $r=0$, while for the even pair $(2,3)$ it has a  local minimum.}
\end{figure*}  

Quantum discord is defined as minimum difference between mutual informations expressions, $J(\rho_{12})$ and $I_{\phi',
\theta'}(\rho_{12})$ $^{13} $, for composite state of site 1 and 2. 
 Now, the conditional density matrix of qubit 1, conditioned on the measurement of 2. is given as
\begin{equation}
 \rho_{1|2_{\tilde{k}}}=\frac{1}{p_{\tilde{k}}}
Tr_2
\ket{\tilde{k}} \bra{\tilde{k}} \rho_{12}.
 \end{equation} 
 Both the mutual information and the conditional entropy now depend on the measurement basis. We express the mutual information which depends on two parameters of the measurement basis as
 \begin{equation}
 I_{\phi^\prime,\theta'}(\rho_{1,2})= S(\rho_{1})- \mathcal{S}_{\phi',\theta'}(\rho_{(1|2)}).
 \end{equation}
 Here the conditional entropy is defined as the weighted sum of the marginal conditional entropies as,
 \begin{equation}
 \mathcal{S}_{(\phi',\theta')}(\rho_{1|2})= p_{\tilde{0}}S(\rho_{1|\tilde{0}})+p_{\tilde{1}}S(\rho_{1|\tilde{1}}).
 \end{equation}
The minimum difference of the two mutual information expressions over all range of $\phi$ and $\theta'$ is defined as quantum discord, we have
\begin{equation}
D{(1,2)} = min\left(\mathcal{S}_{\phi,\theta'}\left(\rho_{1|2}\right)\right) - S(\rho_{(1,2)}) + S(\rho_1)).
\end{equation}
 For minimising the  conditional entropy, we first look at eigenvalues of the conditional density matrix $\rho_{1|2_{\tilde{k}}}$. From the two-qubit density matrix
 $\rho_{1,2}$ given in Eq.12, we can calculate the conditional reduced density matrix, we have 
 \begin{equation}
 \rho_{1|2_{\tilde{0}}}=
  \frac{1}{p_{\tilde{0}}}
  \begin{pmatrix}
    {\cos^2 \frac{\theta'}{2}}u + \sin^2\frac{\theta'}{2} w & z^*  \\
    z & \sin^2\frac{\theta'}{2} v + {\cos^2 \frac{\theta'}{2}}w \\ 
  \end{pmatrix},
\end{equation}  
where $z=e^{-i\phi}x \sin\frac{\theta'}{2}\cos \frac{\theta'}{2}$, the normalisation $ p_{\tilde{0}}=u\cos^2(\theta'/2)+v\sin^2(\theta'/2)+w$.  The other conditional reduced matrix 
$\rho_{1/2_{\tilde 1} }$ is obtained by replacing $\theta^\prime$ by $\theta^\prime +\pi$,  and the azimuthal angle $\phi^\prime$ by $-\phi^\prime$. The eigenvalues of the above matrix are straightforward, and from the eigenvalues we can calculate the von Neumann entropy\cite{kundu}. It turns out that the minimum for the conditional entropy $\mathcal S_{\phi^\prime,\theta^\prime}$ occurs for $\phi^\prime=0,\theta^\prime=\pi/2$, for all parameter values in our case, unlike the anisotropic Heisenberg  model ground state. The minimum conditional entropy can be expressed as
a Shannon binary entropy function, $H(p)=-p\log{p}-(1-p)\log{1-p}$. Thus, we have,
 
\begin{equation}
\mathcal{S}_{0,\frac{\pi}{2}}(\rho_{1|2})= H\left( \frac{1+\sqrt{(u-v)^2+4x^2}}{2}\right) .
\end{equation} 
Using the minimised conditional entropy, we get a simple expression for the quantum discord, we have
\begin{equation}
D{(1,2)} = \mathcal{S}_{0,\frac{\pi}{2}}(\rho_{1|2}) - S(\rho_{1,2}) + S(\rho_1),
\end{equation}
in terms of the conditional entropy, the von Neumann entropies of the two-qubit density matrix and the reduced density matrix of the first spin. The von Neumann entropy of a density matrix
$\rho$  in terms of its eigenvalues $\lambda_i$  is given by $S(\rho)=-Tr \rho \log {\rho}=-\sum \lambda_i \log{\lambda_i}$.
The eigenvalues of the composite density matrix $\rho_{1,2}$ are  $w,w, (u+v\pm\sqrt{(u-v)^2+4x^2})/2$, 
 and the eigenvalues for the one-qubit density matrix $\rho_{1}$ are $u+w$ and $v+w$. 

  The discord has many contributions from different entropies appearing in Eq.35. The minimised conditional entropy is shown in Fig.4 as a function of $r$, the ratio of the interaction strengths, for a chain of $N=100$ spins for a few values of the magnetic field. The inset shows it as a function of the magnetic field, for a few different values of $r$.   
We can see from Fig.4, that the conditional entropy always increases as $r$ increases and the curve at $h/j_x=0$ lies above the curves for non-zero $h/j_x$ values, which explains the behaviour of the entropy exhibiting a peak as a function of $h/j_x$ shown in the inset, at zero magnetic field. We can see from the inset of Fig.4 that the maximum of the conditional entropy occurs at $r=1$. This can be understood from the fact that it has a sharp increase in $r\simeq 1$ region, that is more apparent from $h/j_x=0$ curve shown in the main figure. The behaviour of minimised conditional entropy is similar to the diagonal correlation function $w_{odd}$. This is as expected as the conditional entropy is just a Shannon binary entropy,
shown in Eq.34. The Shannon function $H(p)$ and the product $p(1-p)= n_1(1-n_1)+x_{odd}^2$ have similar behaviour, The diagonal correlation function $w_{odd}$ has a similar expression, with the difference that $x_{odd}^2$ appears with a negative sign. 
Now, the other additive contribution for the discord comes from the von Neumann entropy of the single-qubit reduce density matrix, $S(\rho_1)$, which increases the discord by an amount
$n_1(1-n_1)$. The entropy of the composite density matrix is always a reducing factor. The discords for the odd pair (1,2) and the even pair (2,3) have been plotted in Fig.5(a) as a function of $r$ at different $h/j_x$-values. The behaviour of the discord is similar to the  discord in the ground state of the transverse-Ising model and XYZ spin model,  where it is maximal near the critical point$^{[15]}$. At $r=0$, $D(1,2)$ has its maximum value but it decreases rapidly in the regime $0<r<1$  which reflects that the entropy contribution from reduce density matrix $S(\rho_1)$ that dominates in deciding the behaviour of the discord. But in the regime $r>1$, the conditional entropy and $S(\rho_1)$ both grow but the discord still decreases because of the growth of composite entropy in this region. As is expected, it can be seen from comparing Fig.1(b) and Fig.5(a) that the quantum discord starts emerging early than concurrence for the even pair $(2,3)$, and does not fall as quick as the concurrence for the odd pair $(1,2)$ which indicates the presence of the quantum correlation even when the ground state is separable in those regions. The discord as a function of ratio of magnetic field and interaction strength, $h/j_x$ is shown in Fig.5(b) where we see that for the odd pair, $D(1,2)$ has maximum at $h=0$ and falls off quickly as $h$ increases while for the even pair, $D(2,3)$ has its local minima at the critical point.  This pattern is similar to the pattern of the concurrence, as shown in Fig.1(a), with a difference that at an arbitrary $r$-value the concurrence is higher than the discord, except for $h/j_x \simeq 0$ where both approach to 1. For $r=1$, the behaviour of the concurrence and the discord are different in the critical region. The discord approaches its maximum value while the concurrence shows a dip giving a local minimum at the critical point, $h=0$. This difference in the behaviour of the concurrence and the discord arises from the significant contribution of the entropy of the reduce density matrix $S(\rho_1)$.

\section{\label{sec:level5}Global Entanglement:}
Correlation measures we discussed in the previous sections,  the concurrence and the quantum discord,  focused primarily bipartite entanglement and quantum correlations. The analytical approach works well only for nearest-neighbour pairs of qubits. Going beyond neighbouring spins, the correlation functions are difficult to calculate analytically, as various phase factors pertaining to the intermediate qubits appear, making it difficult to go beyond nearest neighbour pairs. For instance, the analogue of Eq.19 for  the correlation function of qubits 1 and 3 will involve operators pertaining to qubit 2 also.
The nearest-neighbour concurrence and discord computed and discussed above
do not count give an insight into the multi-partite entanglement structure of the spin chain. We will investigate the global entanglement measure in this section, and a multi-species entanglement in the next section to discuss the multi-party quantum correlations in the ground state of the Kitaev spin chain. 

Mayer and Wallach\cite{meyer} proposed a scalar measure of pure state entanglement in three and four qubit states. It is defined as n-particle entanglement as it considers entanglement of each site with the rest of the system. It vanishes if the n-particle state is completely separable. Brennen in his paper\cite{brennen} generalized and derived it for any number of qubits in the form :\\
\begin{equation}
 E_{global} = \frac{2}{N}\sum\limits_{i=1}^N \left(1-Tr(\rho_i^2)\right) .
\end{equation}
where $\rho_i$ the reduced density matrix of $i'$th qubit,
Thus, the global entanglement for a multi-particle state is expressed as the average of the entanglement of each site to the rest of the system. As,   $Tr(\rho_i^2)$  is a measure of purity of the local reduced density matrix,  the global entanglement has a physical meaning of being the average qubit purity of the multi-qubit state. It is the averaging procedure that makes it a global measure.. 

Though there are two inequivalent nearest-neighbour pairs, the odd and even pairs, for $J_x\ne J_y$ in the Kitaev chain, all the qubits are identical in the chain, i.e. the one-qubit reduced density matrix is the same for all qubits, as shown in Eq. 10. 
The expression for the global entanglement can be simplified in terms of the local average of the fermion occupation number, we have
\begin{equation}
E_{global}=4\langle c_1^\dagger c_1\rangle(1- \langle c_1^\dagger c_1\rangle).
\end{equation} 
The average fermion occupation can be calculated using Eq.11.

The global entanglement is plotted in Fig.6(a), for a chain of $N=100$ spins as a function of the magnetic field for a few different values of $r$, the ratio of the interaction strengths, and as a function of $r$  for a few values of the magnetic field in Fig.6(b). For a constant strength ratio $r$, the global entanglement achieves its maximum for $h=0$ and decreases as magnetic field increases. The first order derivative has been shown in inset of Fig. 6(a), it shows a smooth behaviour near the critical point similar to the concurrence, thus it can only give a weak signal of the quantum phase transition.

 \begin{figure*}[]
 {(a) \hskip 6.5 cm (b)}\\
\includegraphics[width=0.49\linewidth, height=6.5cm]{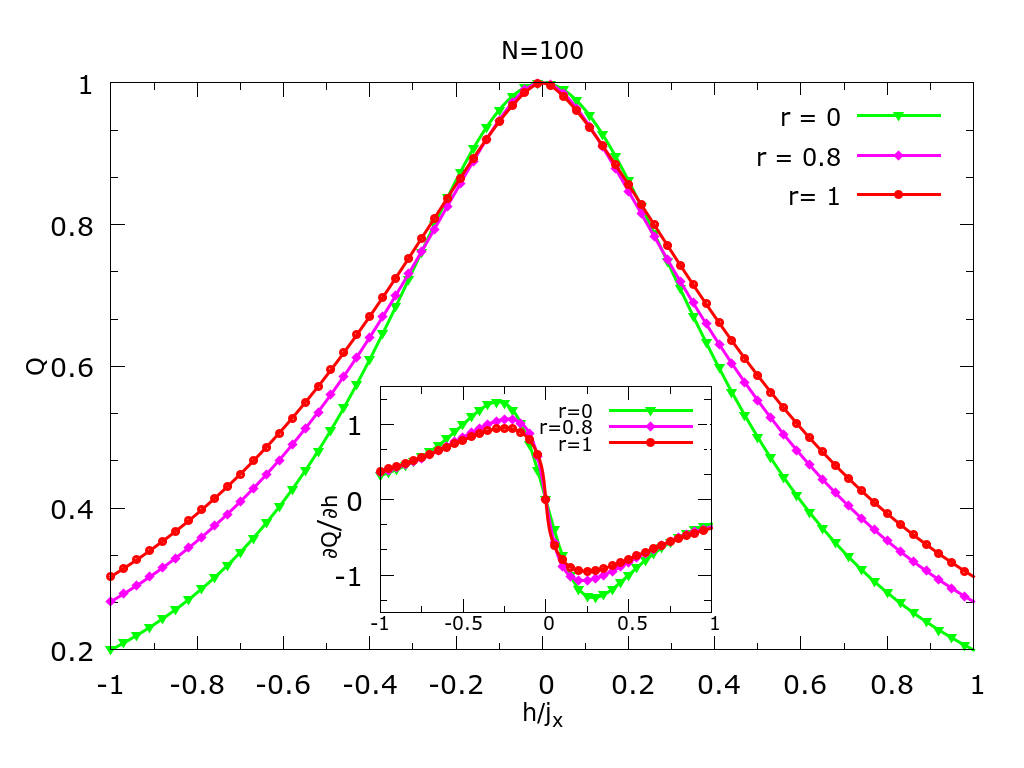} 
 \includegraphics[width=0.49\linewidth, height=6.5cm]{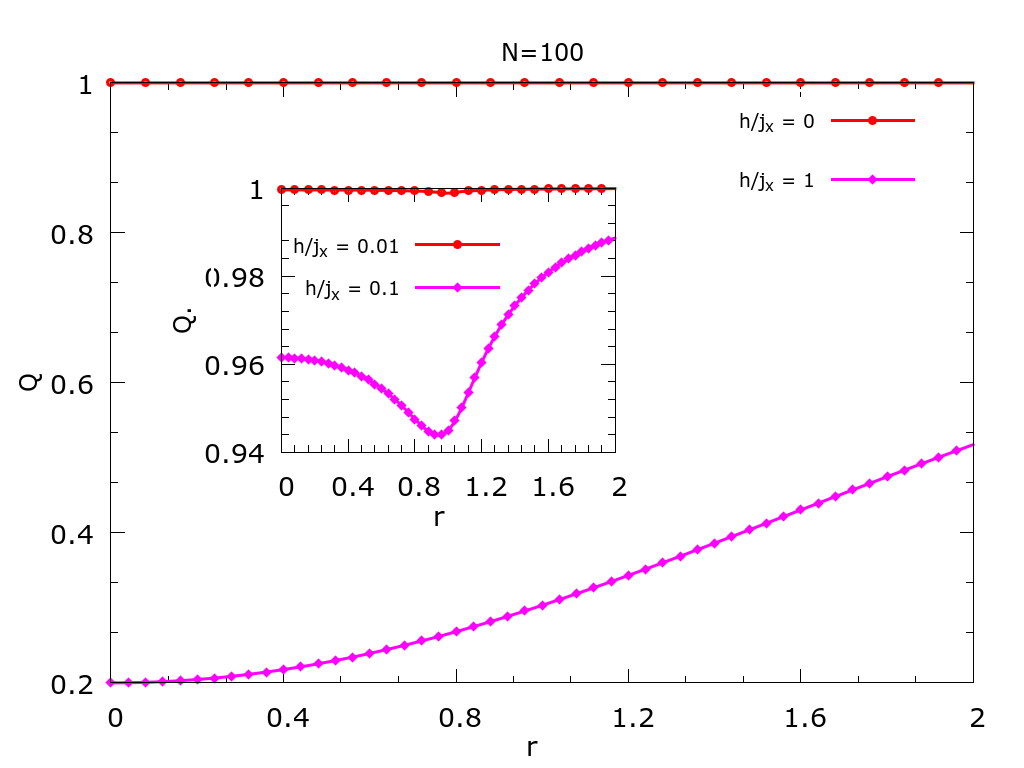}
\caption{(a). The global Entanglement as a function of $h/j_x$ at different strength ratio parameter, $r$. At any $r$-value it approaches to unity at $h=0$. This is in contrast with the other entanglement measures. The  first-order derivatives is shown in the inset as a function of $h/j_x$. (b). The global entanglement  as a function of $r$, for different values of magnetic field. The inset plot shows very quick variation of Q when magnetic field is very small.}
\label{fig:image2}
\end{figure*}

 In constant magnetic field, the entanglement increases smoothly and tends towards unity as the strength ratio $r\rightarrow \infty$. For $h=0$ the number operator term is constant, and therefore, the sum over all modes gives a normalised value of unity (see Fig.6(b)). The inset plot shows the global entanglement  near the critical region, $h\rightarrow 0$. exhibiting an anomalous dip near comparable interaction strength $j_x\simeq j_y$.  But, this dip reduces and eventually disappears as $h$ increases beyond $h>0.1$, as can be seen from the main part of Fig.6(b).
In the case of $j_y=0$, the Hamiltonian can be solved for just $N=2$, as the system breaks into $N/2$ disconnected pairs. The lowest eigenvalue is given by $\lambda=-\sqrt{j_x^2+4h^2}$ and the ground state is given by $ a\Ket{00}+ b\Ket{11}$.
Here, for $j_x=1$, the amplitude coefficients are  $a=1/\sqrt{(\lambda-2h)^2}$ and $b=\sqrt{1-a^2}$. Thus, the global entanglement is given by $4|a|^2(1-|a|^2)$,  and for this case it takes the value $Q=0.2$. Its behaviour of approaching unity at any $r$-value (see Fig.6(a)) can be explained by Eq.22, the expression for the concurrence, the absence of the off-diagonal correlation function in Eq.37 causes the global entanglement rise to unity at any $r$ value. The smooth behaviour of the derivatives can also be explained using a similar argument. Thus the global entanglement does not show any singular behaviour as a signature in the vicinity of a quantum critical point.

\section{\label{sec:level6}Macroscopic Multi-Species Entanglement:}
In the previous sections, we have discussed quantum correlations between sub systems of a large system, viz. entanglement and discord of two qubits, and the entanglement of a qubit
with the rest of the system. All these pertain to the correlations between two spatial partitions. In this section, 
we will focus on the entanglement between two species of particles, viz. spin up and spin down particles, that are spread over the whole system. This aspect has been explored, and  has been shown to track quantum phase transitions\cite{subra2} in the context of transverse-field Ising model. We will investigate, in this section, the entanglement between the up and down spins in the ground state of the Kitaev spin chain.

We can view a spin state of $N$ spins as a state of two species of particles, with $N_\uparrow$ up-spin particles and  $N_\downarrow$ down-spin particles, occupying $N$ sites of the lattice, with a hard-core repulsion, that forbids double occupancy. Further, we have a condition that the number of particles is equal to the total number of sites, $N=N_\uparrow+N_\downarrow$. Thus, each site has a two-dimensional Hilbert space, namely, a up-spin occupied site and a down-spin occupied site.
An arbitrary pure state of the composite system can be written as  
\begin{equation}
\ket{\psi}= \Sigma_{u,v} ~\phi(u,v) \Ket{u}_\uparrow\Ket{v}_\downarrow , 
\end{equation}
where $\Ket{u}_\uparrow$, $\Ket{u}_\downarrow$ are individual basis states of $\uparrow$ and $\downarrow$ type of particles respectively and $u(v)$ labels the set of sites occupied by $\uparrow (\downarrow)$ type of particles. 
Here $\phi(u,v)$ is the wave function amplitude determining all properties of the state. A  non product wave function will imply an entanglement between the two species, that can
be measured by the von Neumann entropy of  $up-spin$ particles (or equivalently the entropy of the down-spin particles). This entropy is a thermodynamic quantity, as it will be proportional to the system size, as both the types of particles occupy the full size of the system.  Further we can also use $u$, the set of locations of the up-spin particles, to label the locations where down-spin locations, as the complement of $v$ (which is just $u$) serves as good as $v$. Thus, the above state will take an explicitly Schmidt-decomposed form, we have
 \begin{equation}
  \ket{\psi}= \Sigma_{u} ~\phi(u) \Ket{u}_\uparrow\Ket{u}_\downarrow,
\end{equation}
where the Schmidt number  is just the probability $|\phi(u)|^2$. It is straightforward to get the reduced density of up-spin particles, we have  
\begin{equation}
\rho_\uparrow= \Sigma_{u} ~|\phi(u)|^2 \Ket{u}_\uparrow \langle{u}|_\uparrow.
\end{equation}
The von Neumann entropy of the reduced density matrix $\rho_\uparrow$ is given by  $S_\uparrow=\Sigma_{u} ~|\phi(u)|^2 \log |\phi(u)|^2$.

\begin{figure*}[t]
\includegraphics[height=15cm,width=18.5cm]{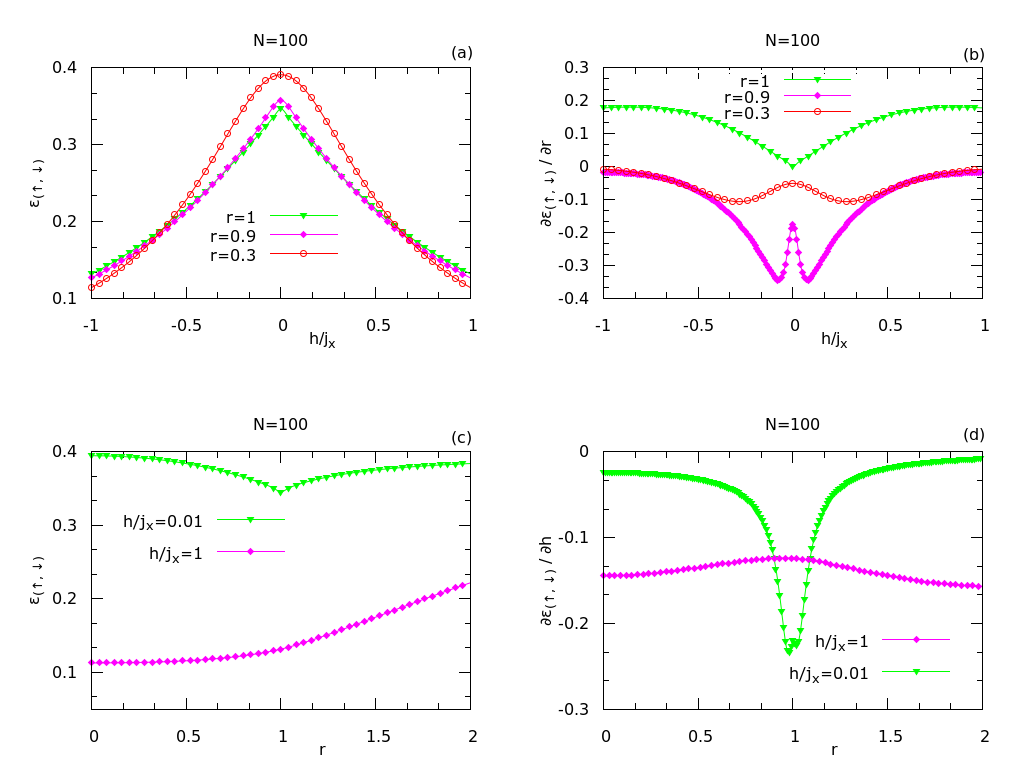} 
 
\caption{(a). The multi-species entanglement as a function of $h/j_x$, for different $r$ values. (b) The first derivate with respect to $r$ is shown as a function of $h/j_x$ for
different $r$ values. (c)  The multi-species entanglement is shown as a function of $r$ for different $h$. (d) The first derivative with respect to  the magnetic field is shown  as a function of
$r$, for different $h/j_x$ values.  The total number of sites is $N=100$ in all the cases, This measure of entanglement, among all the measures studied, shows the best signature of the critical behaviour, with a pronounced peak and dip as seen in the figures. }
\label{fig:image2}
\end{figure*}

The ground state, shown in Eq.7, of the Kitaev chain is given as a direct product over $q$  in the occupation number basis of the mode operators. For each $q$, the state is a linear
combination of the six basis states shown. The first basis state that appears is $|1001\rangle= |\uparrow_{q-\pi},\downarrow_{-q},\downarrow_{q},\uparrow_{\pi-q}\rangle$, which denotes the occupation of $q-\pi$ and $\pi-q$ modes ($n_{q-\pi}=n_{\pi-q}=1$), and the other two modes are unoccupied ($n_{-q}=n_q=0$). On the other hand, the occupied mode implies the presence of $\uparrow$, and the unoccupied mode implies the presence of $\downarrow$ particle. That is, certain number of modes are associated with $\uparrow $ and the rest of the modes are associated with $\downarrow$ occupation.  This can be also be stated for a given basis state, a set of modes are associated with $\uparrow$ occupation, and the same set is associated with $\downarrow$ un-occupation.  Thus, we can view the six basis states, of Eq.7, from the view-point of two-species occupation as direct products of $\uparrow$ and $\downarrow$ states, we have
\begin{eqnarray}
|1001\rangle &\equiv& |q-\pi,\pi-q\rangle_{\uparrow} |q-\pi,\pi-q\rangle_{\downarrow}, \nonumber \cr
|1111\rangle  &\equiv &|q-\pi,-q,q,\pi-q\rangle_{\uparrow} |q-\pi,-q,q,\pi-q\rangle_{\downarrow},\cr
|1100\rangle &\equiv& |q-\pi,-q\rangle_{\uparrow} |q-\pi,-q\rangle_{\downarrow}, \nonumber \cr
|0000\rangle &\equiv& |o\rangle_{\uparrow} |o\rangle_{\downarrow}, \nonumber \cr
|0011\rangle &\equiv& |q,\pi-q\rangle_{\uparrow} |q,\pi-q\rangle_{\downarrow}, \nonumber \cr
|0110\rangle &\equiv& |-q,q\rangle_{\uparrow} |-q,q\rangle_{\downarrow}.
\end{eqnarray}
 We can rewrite the ground state as linear combination of product states of $\uparrow$ and $\downarrow$ particle occupation, as
\begin{eqnarray}
& \ket{GS}&= \prod_q  \alpha \Ket{ad}_\uparrow\Ket{ad}_\downarrow+ \beta \Ket{abcd}_\uparrow\Ket{abcd}_\downarrow+ \gamma
 \Ket{ab}_\uparrow\Ket{ab}_\downarrow \cr &+ &
  (1-\beta) \Ket{0}_\uparrow\Ket{0}_\downarrow
   -\gamma \Ket{cd}_\uparrow\Ket{cd}_\downarrow+ \alpha \Ket{bc}_\uparrow \Ket{bc}_\downarrow,
\end{eqnarray}
where the labels $a,b,c,d$ are used for the four different momentum values of a mode, respectively for $q-\pi,-q,q,\pi-q$. In each component,  the state for $\uparrow$ is denoted by the occupied state for the momentum value, and the state for $\downarrow$  is denoted by the unoccupied state of the momentum value. The state $|0\rangle_\uparrow |0\rangle_\downarrow$ has no occupation of $\uparrow$ and no un-occupation for $\downarrow$ for all the four momentum values of the mode. Since the above state is all ready in a Schmidt-decomposed form, the reduced density matrix for $\uparrow$ particles is straightforwardly evaluated by a partial trace over the states of $\downarrow$ particles,
 $\rho_{\uparrow} =Tr_{\downarrow}|GS\rangle\langle GS|$. The eigenvalues of $\rho_\uparrow$ can be written as a product over $q$ of $ \Lambda_q (n_a,n_b,n_c,n_d)$, where $n_{i}$ is either 0 and 1,
 depending on whether the mode $i$ is unoccupied or occupied respectively for the momentum $q$. For instance, $\Lambda_q(1,0,0,1)=\alpha(q)$ is the first coefficient in the ground state
 of Kitaev chain, and similarly the other coefficients will appear for different values of $n_i$. A typical eigenvalue of the reduced density matrix is given by,
 \begin{equation}
  \lambda= \prod_{0\leq q \leq \pi/2}^{}\Lambda_{q}(n_{a},n_{b},n_{c},n_{d}).
  \end{equation}
  There are $2^4$ possible values for each mode $q$, and there are $N/4$ modes, giving $2^N$ eigenvalues of $\rho_\uparrow$.
  The entropy of the subsystem defined as $S_\uparrow =-\Sigma \lambda\log \lambda$, can be further simplified in terms of a sum over the modes, we have
\begin{equation}
S_{\uparrow} = -\sum_{0\leq q \leq \pi/2}\sum_{i=1}^6 |x_{i}|^2\log |x_{i}|^2.
\end{equation}
Here $x_{i}$ is one of 6 wave function amplitude of the ground state wave function shown above. 
The wave function is symmetric with respect to up and down spins. That is, a partial trace over either up-spin states or down-spin states gives the same reduced density matrix, implying that the  entropy for up spins is the same as the entropy for down spins. This entropy is taken as the entanglement between the up and down spins, as it is usually taken as the measure of the bipartite entanglement in a pure state. Unlike, the case of bipartite entanglement of spatial partitions, here both up spins and down spins access the full physical size of the system.
Thus, the above entropy is proportional to the total number of sites $N$, we have a macroscopic thermodynamics entropy that grows with the system size $N$.  We define the macroscopic entanglement in the thermodynamic limit as 
\begin{equation}
\epsilon_{\uparrow,\downarrow}= {\rm Lim} {N\rightarrow\infty}\frac{S_\uparrow}{N}.
\end{equation}
We expect that the multi-species entanglement, as characterised by the macroscopic von Neumann entropy of the reduced density matrix for the up spins, will exhibit singular behaviour
close to the critical point at zero magnetic field.

In fig.7(a), we show the multi-species entanglement $\epsilon_{\uparrow,\downarrow}$ as a function of the magnetic field $h/j_x$ for $N=100$. Similar to the other correlation measures we discussed in the previous sections, the entanglement between up and down spins reaches to its maximum at critical point $h=0$ and as magnetic field is increased the entanglement decreases for a fix $r$-value. As the $r-$value is decreased, the entanglement increases in the critical region, but below $r=0.3$, it does not increase further (the curve for this regime is not shown in the figure). 
The entanglement for a larger field of $|h/j_x|>0.6$ changes its pattern and is more for larger value of  $r$. The first-order derivative of this entanglement gives local maximum at the critical point, shown in Fig.7(b). In the critical region, for increasing $r$-values, the local maximum gets sharper and more pronounced. But at $r=1$, it suddenly disappears and minimises itself which can be seen as a signal of the quantum phase transition. In Fig.7(c), the entanglement has been plotted as a function of $r$ at different  $h/j_x$-values. For $h\simeq 0$, it remains constant in lower range of $r$ but then it rapidly increases in $r\simeq 1$ region and then saturates at large values of $r$. For $h\rightarrow 0$, the entanglement minimises at equal interaction strengths and reaches to  a saturation value for either larger or smaller range of $r$. The first-order derivative of entanglement with respect to $h$ is shown in Fig.7(d) where it shows a characteristic of a near divergence at $r=1$ for $h\rightarrow 0$, as a signature of the critical point behaviour
\paragraph*{} 
\section{\label{sec:level7} Conclusions}

In conclusion, we have investigated the quantum correlations and entanglement measures in the ground state of a Kitaev-type spin chain, with alternating interaction strength $j_x$ for x-components of the neighbouring spins and the interaction strength $j_y$ for y-components of the neighbouring spins, along with a transverse magnetic filed $h$.
We have analytically computed the concurrence measure of the pair entanglement for odd and even pairs of neighbouring spins.
The odd concurrence approaches to its maximum at the critical point and falls quickly as $h$ increases. The concurrence falls off as a function of the size of the system, as $1/N$ order for small $N$, but converges quickly above $N=40$. The odd concurrence shows a cutoff behaviour as a function of the ratio of the interaction strength. The first-order derivatives of odd concurrence $C(1,2)$ and even concurrence $C(2,3)$ do not show any singular divergence in the critical range $h\rightarrow 0$, in sharp contrast with the singular behaviour of the concurrence in the ground state of the transverse XY model.
It is seen that  the second-order derivatives too exhibit a smooth behaviour in the critical region (which are not discussed here). 

We have studied the quantum discord for neighbouring spins as a function of magnetic field and the ratio of the interaction strengths. The discord is always nonzero for all values of $r$ at any $h$ values, unlike the concurrence that exhibits a cutoff behaviour. Similar to the concurrence, the odd pair discord takes its maximum value in the critical region as $h\rightarrow0$, approaches to unity for $r=0$. The discord and the concurrence behave differently in the critical region  at $r=1$, the discord reaches to a maximum while concurrence reaches its local minimum (see Fig.5(b) and 1(a)). 

We  have further examined a global entanglement measure, which, similar to the concurrence and the discord, touches a maxima at $h=0$, independent of the value of the interaction strengths. The global entanglement is the simplest measure to calculate, as it depends only on single-qubit reduced density matrices, nevertheless it does show similar behaviour seen in
the bipartite entanglement and the quantum discord. Lastly, we investigate the multi-species entanglement, a macroscopic and thermodynamic quantity, between the up spins and the down spins in the ground state Kitaev chain. The multi-species entanglement, like  other correlation measures studied, showed  a peak structure in the critical region. Among all the correlation measures, it exhibits the most-pronounced signature, a singular-like behaviour in the critical region, for the case of $j_x=j_y$.
 The singularity in the first-order derivative with respect to $r$ can be seen as a signal of quantum phase transition in the one dimensional Kitaev model.

\end{document}